\documentclass[journal]{IEEEtran}
\pdfoutput=1
\usepackage{cite}
\ifCLASSINFOpdf
\else
\fi
\newcommand{\field}[1]{\mathbb{#1}}

\newcommand{\R}{\field{R}} 
\newcommand{\commentOut}[1]{}
\newcommand{\Th}{^{\text{th}}}

\usepackage{subcaption}
\usepackage{graphicx}

\usepackage{svg}

\usepackage{amsmath,amssymb,amsthm}

\usepackage{algorithm}
\usepackage{algorithmic}
\usepackage{comment}
\usepackage{float}
\usepackage{array}
\usepackage{makecell}
\usepackage{graphicx}
\usepackage{multirow}
\usepackage{balance}
\makeatletter
\newcommand*\titleheader[1]{\gdef\@titleheader{#1}}
\AtBeginDocument{%
\let\st@red@title\@title
\def\@title{%
\bgroup\normalfont\large\centering\@titleheader\par\egroup
\vskip1.5em\st@red@title}
}
\makeatother

\makeatletter
\def\endthebibliography{%
	\def\@noitemerr{\@latex@warning{Empty `thebibliography' environment}}%
	\endlist
}
\makeatother

\begin{document}
	%
	\title{A Machine Learning Framework for Event Identification via Modal Analysis of PMU Data}
	%
	%
	%
	
	\author{Nima~Taghipourbazargani,~\IEEEmembership{Member,~IEEE,}
	    Gautam~Dasarathy,~\IEEEmembership{Senior~Member,~IEEE,}
		Lalitha~Sankar,~\IEEEmembership{Senior~Member,~IEEE,}
		and~Oliver~Kosut,~\IEEEmembership{Member,~IEEE}
		\thanks{The authors are with the School of Electrical, Computer, and Energy
			Engineering, Arizona State University, Tempe, AZ 85281 USA (e-mail:
			\{ntaghip1,gautamd,lalithasankar,okosut\}@asu.edu).}}

	\markboth{IEEE TRANSACTIONS ON POWER SYSTEMS}%
	{T.Bazargani\MakeLowercase{\textit{et al.}}: A Machine Learning Framework for Event Identification via Modal Analysis of PMU Data}
	
	\maketitle

\begin{abstract}


Power systems are prone to a variety of events (e.g. line trips and generation loss) and real-time identification of such events is crucial in terms of situational awareness, reliability, and security. Using measurements from multiple synchrophasors, i.e., phasor measurement units (PMUs), we propose to identify events by extracting features based on modal dynamics. 
We combine such traditional physics-based feature extraction methods with machine learning to distinguish different event types. Including all measurement channels at each PMU allows exploiting diverse features but also 
requires learning classification models over a high-dimensional space. 
To address this issue, various feature selection methods are implemented to choose the best subset of features.
Using the obtained subset of features, we investigate the performance of two well-known classification models, namely, logistic regression (LR) and support vector machines (SVM) to identify generation loss and line trip events in two datasets. The first dataset is obtained from simulated generation loss and line trip events in the Texas 2000-bus synthetic grid.
The second is a proprietary dataset with labeled events obtained from a large utility in the USA involving measurements from nearly 500 PMUs. 
Our results indicate that the proposed framework is promising for identifying the two types of events.
\end{abstract}



%
\IEEEpeerreviewmaketitle

\section{Introduction}
%
%
%
%


    \IEEEPARstart{G}{iven} the increased penetration of intermittent renewable energy sources (e.g., solar and wind) as well as unconventional loads (e.g. electric vehicles) in the grid, real-time monitoring of system operating conditions has become more vital to ensure system reliability, stability, security, and resilience.  
    Real-time detection and identification of events enhances situational awareness and assists system operators in quickly identifying events and taking suitable remedial control actions to avert disturbances in a timely manner \cite{Partial-Arghandeh}. 

    The problem of event detection---i.e., deciding whether an event has occurred---has been well studied in the literature \cite{Detection, Unsupervised_YangWeng,Detection2}. The problem of event \emph{identification}---deciding which type of event has occured---is more challenging, due to the fact that power systems are inherently nonlinear with complex spatial-temporal dependencies. Thus, it is difficult to develop accurate and sufficiently low order dynamical models that can be used to identify each distinct event \cite{MPinPS}. Hence, in this paper we merely focus on the event identification problem.
    
    Prior research in the context of real-time identification of events in power grids can be categorized into two main approaches, namely model-based and data-driven. 
    Model-based methods (see e.g., \cite{Model-based1,Model-based2,model-based3,model-based5}) involve modeling of power system components and estimation of the system states. The performance of such methods highly depends on the accuracy of dynamic models and estimated states of the system which in turn limits their implementation for real world problems. 

    Data-driven methods have received increased attention, mainly due to the growing penetration of phasor measurement units (PMUs) in the electric grid which can help address situational awareness challenges.
    PMUs provide time-synchronized current and voltage phasor measurements across the grid at high sampling rates thereby allowing operators to capture system dynamics with good precision and fidelity~\cite{Brahma-DynEvents}.

    Data-driven approaches for event identification in the literature can be broadly categorized into unsupervised and supervised approaches. The main difference between the two approaches is that the latter uses labeled data while the former does not. 
        The main disadvantage of unsupervised methods (see, for example, \cite{Unsupervised_YangWeng,Unsupervised1-Ellipsoid,Unsupervised3-Ensemble,Unsupervised4-Kmeans,Unsupervised2-PCA}) is that, although they can distinguish between clusters of events, they do not possess the ground truth to associate each cluster with its real-world meaning. Furthermore, when there is access to even a small amount of labeled data, supervised learning has been shown to perform better than unsupervised learning methods \cite{Unsupervised_YangWeng,Yangweng_limitedlabel}.
        For these reasons, in this paper, we focus on the supervised setting to exploit the available labels in our dataset for the event classification task.

        Available literature in the supervised setting for event identification can be further divided into two subgroups depending on whether they rely on the physics of the system to process the PMU data or not. 1) Physics-based signal processing methods \cite{Brahma-DynEvents,SignalProc-1,SignalProc-2} such as modal analysis for feature extraction can be directly applied to PMU measurements to identify events. The key idea in such approaches,  often referred to as mode decomposition, is to identify system events by thresholding the coefficients of some basis functions. However, due to the diversity of power system events, choosing proper thresholds for different scenarios is not an easy task. Another example of physics-based feature extraction method is \cite{li2018real} where the low-dimensional subspace spanned by the dominant singular vectors of a PMU data matrix is used to characterize an event. The intuition is that similar events would produce data in similar subspaces; the average subspace angle is used to quantify this similarity. Although the proposed approach seems to be promising in identifying different types of events, it requires a large number of labeled events to construct the dictionary of events corresponding to varying operation conditions and different types of events. 
        2) Event identification methods such as \cite{heuristic_1,heuristic_2} extract model-free features from PMU data and classify various types of events based on the similarity of the extracted features. For instance,  \cite{heuristic_1,heuristic_2} use the PMU time series data to construct a minimum volume enclosing ellipsoid (MVEE) and then properties of this ellipsoid such as volume, rate of change of the volume, etc., are used to identify different types of events. However, the high computational time of the such methods limits their efficiency for real-time application \cite{Ellipsoid_time}. Within the same category, machine learning (ML) based methods \cite{supervised1-DT-Vittal,Supervised6-CNN,Supervised2-KNN,Supervised3-SVM,Supervised4-ELM,CNN_new1,TextMining_Sara,DNN_new,LSTM_new} either use time series PMU data or their pruned version (see, for example, \cite{TextMining_Sara}) and feed them into a machine learning classification model to identify various types of events \footnote{Due to the space limitation, we are not covering the details of these approaches.}. 
        The main limitations to these approaches can be summarized as follows: a) such models require a large number of labeled events to construct an effective classification model \cite{Supervised6-CNN, CNN_new1, TextMining_Sara,DNN_new}, b) the learned classifiers may not have clear physical interpretations \cite{Supervised6-CNN, CNN_new1, TextMining_Sara,LSTM_new}, and c) the effectiveness of some of these methods in identifying events from real-world PMU data has not been investigated (e.g., \cite{TextMining_Sara}).%
        

        


    We introduce a framework that combines several of the above ideas by exploiting the knowledge of the physics of the system to extract features, and subsequently apply ML techniques to produce a robust classifier from limited but feature-rich training data. 
    Our key contributions are
\begin{itemize}
     \begin{figure*}[h!]
        \centering
        \includegraphics[trim={0 300 0 0},clip, scale = 0.38]{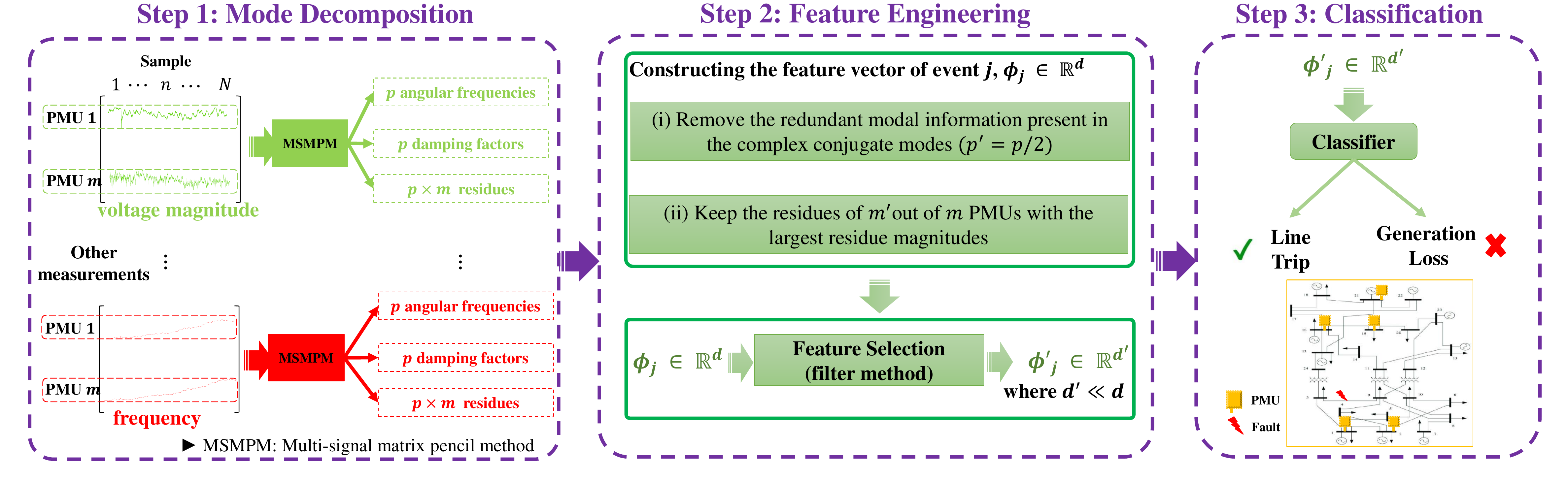}
        \caption{Overview of the proposed event identification framework} 
        \label{fig:overview}
\end{figure*}
    \item Characterizing events based on a set of physically interpretable features (i.e., angular frequencies, damping factors and the residues) obtained from modal analysis of various spatio-temporally correlated PMU measurements. 
    \item Using a well-known approach in machine learning, called bootstrapping, to address the problem of the small number of labeled samples.
    \item Learning a set of robust classification models which can identify generation loss and line trip events in both real and synthetic datasets.
\end{itemize}

An overview of the proposed framework is shown in Fig. \ref{fig:overview}.
In Step 1, using the fact that temporal effects in a power system event are driven by the interacting dynamics of the system components, we use mode decomposition to extract features.
Mode decomposition characterizes each event as a set of features: angular frequencies, damping factors, and the corresponding residues. However, extracting features using all channels of PMU measurements (magnitudes and angles for positive sequence voltages and currents, and frequency) across multiple PMUs will inevitably lead to a high-dimensional feature set, and thus, a key question is to determine which subset of these features can guarantee accurate classification performance.

Step 2 uses filter methods to select features so as to avoid overfitting while ensuring that events can be distinguished by the same set of features. Since filter methods, in contrast with wrapper and embedded methods, are independent from classification models \cite{zebari2020comprehensiveFS}, they are computationally inexpensive and are more efficient for real time applications. 

Finally, in Step 3, a classification model distinguishes generation loss and line trip events based on the extracted features. We test two classification models: logistic regression (LR) and support vector machine (SVM) with radial basis function (RBF).

We use two datasets, one synthetic and one real, to evaluate the performance of our framework.

To highlight the value of using multiple PMU channels and the combination of physics-based and data-driven methods in our proposed approach, we make a comparison with a well established event identification method based on the work of Li et al. \cite{li2018real}. As detailed in Section \ref{sec:simul}, our results on both real and simulated datasets indicate that the proposed framework is promising for identifying the two types of events.


The remainder of the paper is organized as follows. Section \ref{sec: Math} briefly explains modal analysis and MPM. The proposed framework for feature extraction and feature selection from time series PMU data are presented in Section \ref{sec: FE}. Section \ref{sec: valid} describes the validation methodology which is used to evaluate the performance of models based on different feature selection techniques. In Section \ref{sec:simul}, we illustrate the validity of the proposed framework. Finally, Section \ref{sec: concl} concludes the paper. 
\ifCLASSOPTIONcaptionsoff
\newpage
\fi

\section{Problem Setup}\label{sec: Math} 
The first step in identifying a system event from PMU data is to extract the relevant features from the data stream. 
Due to the high sampling rate of the PMU data, one could plug in the raw data into a machine learning model. However, it is advantageous to use a set of delineating features that are likely to contain information regarding the event type (henceforth referred to as event class). Using the fact that temporal effects in a power system are driven by the interacting dynamics of system components, we propose to use mode decomposition as the framework with which to extract features. More specifically, we assume that each PMU data stream after an event consists of a superposition of a small number of dominant dynamic modes. Thus, the features will be the frequency and damping ratio of these modes, as well as the residual coefficients indicating the quantity of each mode present in each data stream.

Several modal analysis techniques such as MPM, Prony analysis and dynamic mode decomposition \cite{MPM_Prony,MPM_DMD} have been proposed in literature. Relying on earlier observations that MPM is more robust to noise relative to the above mentioned methods, we will use MPM as the mode decomposition technique. 
In general, every PMU has multiple measurement channels, including positive sequence voltage magnitude (VPM) and corresponding angle (VPA), positive sequence current magnitude (IPM), and corresponding angle (IPA), and frequency (F). Furthermore, multiple PMUs across the grid capture the dynamic response of the system after an event through different measurement channels. Therefore, for a chosen measurement channel, we will use the MSMPM to obtain one optimum set of mode estimates which can accurately represent the underlying dynamic behavior of the system \cite{MPM-noiseresilient}. 

Oftentimes only a small number of modes are triggered after an event. In a noise-free system, it is fairly easy to extract these modes. However, in a noisy system, there exist many other low energy modes that are more likely related to the minor noise variations and can make the classification of the events harder. To ensure accurate classifiers we use the low rank approximation of the Hankel matrix constructed from PMU measurements which allows (i) reducing the effect of noise on the accuracy of mode estimation, and (ii) extracting a small number of dominant modes from noisy PMU measurements.

So in subsection \ref{sec:Modal}, we briefly explain modal analysis as a method to capture signatures of an event. 
Then we discuss the background and theory behind single signal and multi-signal MPM in subsection \ref{sec:MPM}. Finally, in subsection \ref{sec:Hankel}, we discuss the low rank approximation of the Hankel matrix obtained from PMU measurements to estimate the sufficient number of dominant modes.

\subsection{Modal Representation of PMU Measurements}\label{sec:Modal}

Consider an electric grid with $m$ installed PMUs. Recall that each PMU has multiple channels through which we can obtain different types of measurements relative to the bus where the PMU is installed. For the sake of clarity, we focus on one channel (e.g., VPM). Let $y_i(n)\in\R$, $i=1,\dots,m$, and $n=0,\ldots,N-1$, denote the VPM measurement obtained from $i\Th$ PMU at sample $n$ with a sampling period of $T_s$. Note that we use the detrended PMU measurements prior to any modal analysis to ensure that the identified modes are associated with the oscillations after a contingency rather than the noise in the PMU measurements. The trend in the time series data refers to the change in the mean over time \cite{Detrending}. Let $\mathbf{\Tilde{Y}}^{(i)} = [\tilde{y}_i(0),\dots, \tilde{y}_i(N-1)]^T \in \R^{N}$ represent a time series PMU data corresponding to the $i\Th$ PMU and $X = [x(0), \dots, x(N-1)]^T \in \R^{N}$ represents a vector of sample number. 
In order to obtain the detrended PMU measurements, we first find a simple linear least-squares fit to the time series data by solving the following regression problem: 
\begin{equation}
\begin{split}
    \min_{w_0, w_1} \sum_{n=0}^{N-1} 
    (\tilde{y}_i(n) - \hat{y}_i(n) )^2, \qquad i = 1, \dots, m
\end{split}
\end{equation}
where $\hat{y}_i(n) = w_0 + w_1x(n)$ is the predicted value of each PMU measurement. To obtain the detrended values, denoted as $y_i(n)$, we remove the result of the the linear least-squares fit from the original PMU measurements, i.e., $y_i(n) = \tilde{y}_i(n) - \hat{y}_i(n)$, $n = 0, \dots, N-1$.

We assume that $y_i(n)$ after an event consists of a superposition of $p$ common damped sinusoidal modes as 
\begin{equation}\label{eq:modalrep}
y_i(n) = \sum_{k=1}^{p}  R^{(i)}_k \times (Z_k)^n  + \epsilon_i(n), \quad i=1,\ldots,m
\end{equation}
where $\epsilon_i(n)$ represents the noise in the $i\Th$ PMU measurement and $Z_k$ is the $k\Th$ mode associated with the event. We represent each mode as $Z_k = \exp(\lambda_k T_s)$ where $\lambda_k= \sigma_k \pm j \omega_k$ and $\sigma_k$ and $\omega_k$ are the damping factor and angular frequency of the $k\Th$ mode, respectively. Furthermore, residue $R^{(i)}_k$  corresponding to each mode $k$ and $i\Th$ PMU measurement is defined by its magnitude $|R^{(i)}_k|$ and angle $\theta^{(i)}_k$. 

Our goal is to distinguish between various types of events by finding the common modes that are captured by all the PMUs and best represent the underlying dynamical behavior of the system. Thus, we are interested in finding a single set of modes (i.e., $\{Z_k\}_{k=1}^p$) that capture the dynamical behavior of the all the $m$ PMUs simultaneously rather than a different set of modes for each PMU measurement stream. 
Note that the corresponding residue of each mode will be distinct for each PMU measurement. 


Let $\mathbf{Y}^{(i)} = [y_i(0),\ldots,y_i(N-1)]^T \in \R^{N}$, $\mathbf{R}^{(i)} = [R^{(i)}_1,...,R^{(i)}_p]^T \in \R^{p}$, and $\mathbf{Z} = [Z_1,\ldots,Z_p]^T$. We define $\mathcal{V}_{\mathbf{Z}}(N)\in \R^{p \times N}$ as the Vandermonde matrix of the modes, $\mathbf{Z}$, as 
\begin{equation}
 \mathcal{V}_{\mathbf{Z}}(N)=  \underbrace{\begin{bmatrix}
1 & Z_1 & \cdots & Z_1^{N-1} \\
1 & Z_2 & \cdots & Z_2^{N-1} \\
\vdots  & \vdots  & \ddots & \vdots  \\
1 & Z_p & \cdots & Z_p^{N-1} 
\end{bmatrix}}_{p\times N}
\end{equation}
Then \eqref{eq:modalrep}, in the absence of noise can be written in  compact form as 
\begin{equation}\label{eq:Rvalues}
\mathcal{V}_{\mathbf{Z}}(N)^T  \mathbf{R}^{(i)} = \mathbf{Y}^{(i)} 
\end{equation}
Once the modes, $\mathbf{Z}$, are estimated, the corresponding residues $\mathbf{R}^{(i)}$ for each PMU measurement stream, $i=1,\dots,m$, can be obtained by solving \eqref{eq:Rvalues}.

\subsection{Multi-Signal Matrix Pencil Method}\label{sec:MPM}

As mentioned earlier, considering the robustness of MPM against noise, it will be used as the main tool to estimate the parameters of \eqref{eq:modalrep}. 
The MPM involves constructing the Hankel matrix over a block of $N$ samples obtained from the $i\Th$ PMU as
\begin{equation}\label{eq:H_i}
\mathcal{H}_i =\underbrace{\begin{bmatrix}
y_i(0) & y_i(1) & \cdots & y_i(L) \\
y_i(1) & y_i(2) & \cdots & y_i(L+1) \\
\vdots  & \vdots  & \ddots & \vdots  \\
y_i(N-L-1) & y_i(N-L) & \cdots & y_i(N-1)
\end{bmatrix}}_{(N-L)\times (L+1)}
\end{equation}
where $L$ is the pencil parameter. We choose $L=N/2$, since it is known that this will result in the best performance of the MPM in a noisy environment (i.e., the attainment of a variance close to the Cramer-Rao bound)\cite{MPM-noiseresilient}.

Using \eqref{eq:H_i}, let $\mathcal{H}_i^{(1)}$ and $\mathcal{H}_i^{(2)}$ be the matrices consisting of the first and last $L$ columns of $\mathcal{H}_i$, respectively. In a noise free setting , as a consequence of \eqref{eq:modalrep}, we can write  $\mathcal{H}_i^{(1)}$ and  $\mathcal{H}_i^{(2)}$ as
\begin{subequations}\label{eq:H_i12}
\begin{align}
\mathcal{H}_i^{(1)} &= \mathcal{V}_{\mathbf{Z}}(N-L)^T  \mathbf{R}^{(i)}_{D} \mathcal{V}_{\mathbf{Z}}(L)\\
\mathcal{H}_i^{(2)} &= \mathcal{V}_{\mathbf{Z}}(N-L)^T \mathbf{R}^{(i)}_{D} \mathbf{Z}_{D} \mathcal{V}_{\mathbf{Z}}(L)
\end{align}
\end{subequations}
where
\begin{equation}\label{eq:Z_D}
\mathbf{Z}_{D} = \text{diag}(Z_1,Z_2,...,Z_p),
\end{equation}
\begin{equation}
\mathbf{R}^{(i)}_{D} = \text{diag}(R^{(i)}_1,R^{(i)}_2,...,R^{(i)}_p).
\end{equation}

Then, the matrix pencil is defined as 
\begin{equation}\label{eq:MatrixPencil}
\mathcal{H}_i^{(2)}- \lambda \mathcal{H}_i^{(1)} = \mathcal{V}_{\mathbf{Z}}(N-L)^T \mathbf{R}^{(i)}_{D}  (\mathbf{Z}_{D}-\lambda \mathbf{I})\mathcal{V}_{\mathbf{Z}}(L)
\end{equation}
where $\mathbf{I} \in \R^{p\times p}$ is the identity matrix. Note that both $\mathbf{R}^{(i)}_{D}$ and $\mathbf{Z}_{D}$ are full rank $p\times p$ diagonal matrices. Further, if $p<L<N-p$ each of $\mathcal{V}_{\mathbf{Z}}(N-L)^T$ and $\mathcal{V}_{\mathbf{Z}}(L)$ has rank $p$. Then, from \eqref{eq:MatrixPencil}, if $\lambda = Z_k$  for any $k=1,\ldots,p$, then the $k\Th$ row of $\mathbf{Z}_{D}-\lambda \mathbf{I}$ becomes zero, and hence the rank of $\mathcal{H}_i^{(2)}- \lambda \mathcal{H}_i^{(1)}$ is reduced by one. Therefore, the parameters $\{Z_k\}_{k=1}^p$ are exactly the values of $\lambda$ where $\mathcal{H}_i^{(2)}- \lambda \mathcal{H}_i^{(1)}$ has a reduced rank\cite{trinh2019iterative}. This is equivalent to the generalized eigenvalues of the pair $(\mathcal{H}_i^{(2)},\mathcal{H}_i^{(1)})$.


The matrix pencil method described above, which focuses on the measurements obtained from a single PMU, may be extended to find a single set of modes which best represent the underlying dynamical behavior of a set of measurements obtained from multiple PMUs. This is done by vertically concatenating Hankel matrices $\mathcal{H}_1,\dots,\mathcal{H}_m$ corresponding to each PMU measurements over a block of $N$ samples as 
\begin{equation}\label{eq:H2}
\mathbf{H} = \underbrace{
    \begin{bmatrix}
    \mathcal{H}_1\\
    \vdots \\
    \mathcal{H}_i\\
    \vdots \\
    \mathcal{H}_m
    \end{bmatrix}}_{m(N-L)\times (L+1)}
\end{equation}
and the same method which is used for a single measurement stream (see \eqref{eq:H_i12} to \eqref{eq:MatrixPencil}) is applied to the matrix $\mathbf{H}$ to identify a set of modes $\{Z_k\}_{k=1}^p$. Finally, we find the residues corresponding to each mode $k$ and $i\Th$ PMU measurements by solving \eqref{eq:Rvalues}. 

\subsection{Model order approximation}\label{sec:Hankel}

Following the assumption that PMU measurements after an event can be represented as a superposition of $p$ dynamic modes and considering the fact that only a small number of modes are enough to represent the underlying dynamical behavior of the system  ($p \ll L$), one can show that $\text{rank}(\mathbf{H}) = p$  for noise free PMU measurements \cite{zhang2018multichannel}. 
However, in practice PMU measurements are noisy and $\text{rank}(\mathbf{H}) > p$. In this case, for a given $p$, we can partly eliminate the noise by using the singular value decomposition (SVD) to find the rank $p$ approximation of $\mathbf{H}$, denoted as $\mathbf{H}_p$. The approximation $\mathbf{H}_p$ results from keeping the $p$ largest singular values of $\mathbf{H}$ (the remaining singular values are replaced by zero). Using $\mathbf{H}_p$ in MPM also provides minimum variance in the estimation of modes in noise-contaminated PMU measurements (we refer readers to \cite{MPM-Noisy,SARKAR-modelorder} for a comprehensive study of MPM performance in the presence of noise in the PMU measurements).

In practice, however, the parameter $p$ is not known. A reliable way to approximate $p$ in \eqref{eq:modalrep} is to find the best $p$ over all the events in our dataset.  
To this end, we define the rank $p$ approximation error of $\mathbf{H}$ as     
\begin{equation}\label{eq:E_p}
    E_p = \frac{\lVert \mathbf{H} - \mathbf{H}_p \lVert_F}{\lVert \mathbf{H} \rVert_F}
\end{equation}
where $\lVert \mathbf{H} \rVert_F$ is the Frobenius norm of  the matrix $\mathbf{H}$. Furthermore, to verify that the estimated value of parameter $p$ is sufficient for capturing the underlying dynamics of the system, we evaluate the reconstruction error of each PMU measurements, denoted as $E_i$, $i=1,\ldots,m$, as
\begin{equation}\label{eq:E_i}
    E_i= \frac{\|\hat{\mathbf{Y}}^{(i)}-\mathbf{Y}^{(i)}\|}{\|\mathbf{Y}^{(i)}\|} 
\end{equation}
where $\mathbf{Y}^{(i)}$ is the original measurement stream and $\hat{\mathbf{Y}}^{(i)}$ is the reconstructed one based on the mode decomposition. 

Using the equations \eqref{eq:E_p} and \eqref{eq:E_i}, the value of the parameter $p$ is determined such that it ensures both $E_p$ and $E_i$ (obtained from various PMU channels) are less than a predefined threshold for all the events in the dataset. Throughout the paper, we consider that this threshold is $1\%$.  

\section{Feature Engineering of PMU Time Series Data}\label{sec: FE}

To characterize the dynamic response of the power system after an event, modal analysis is conducted on each PMU channel (i.e., VPM, VPA, IPM, IPA, and F) obtained from multiple locations across the grid. For instance, using VPM channel measurements from $m$ PMUs, we obtain a set of features consisting of $p$ angular frequencies, $p$ damping factors and the corresponding magnitude and angle of the residues for each of the $m$ PMUs and $p$ modes. Although mode decomposition is meant to focus on only the physically meaningful features of the dataset, there are still simply too many of them ($m\approx 500$ and $p=6$ in our dataset). To avoid overfitting while ensuring that multiple events can be distinguished by the same set of features, a necessary pre-processing step is to select relevant and most informative features. To this end, we propose a two-step approach to reduce the features into a more manageable number. 
In the first step, we select a subset of features by removing the redundant modal information present in the complex conjugate modes and eliminating the smallest residue magnitude to construct a vector of features that characterizes the dynamic response of the system after an event. The second step is to select the most informative and relevant features using a filter method. The details are provided in the following subsections. 

\subsection{Constructing the feature vector}\label{sec:Pre-Proc}

As discussed in Section \ref{sec: Math}, parameter $p$ represents the number of dominant modes in the PMU data streams and can be obtained by finding the best rank $p$ approximation of $\mathbf{H}$.  
In general, these modes can be real or complex conjugate pairs. In our dataset, typically these modes include only complex conjugate pairs and no real modes (i.e., $p/2$ complex conjugate pairs, yielding $p$ modes in total). In order to remove redundant modal information present in the complex conjugate modes, we only keep one mode from each pair. Thus, we keep only $p'=p/2$ modes in the feature vector for each event.  
However, for a small portion of the events, modal analysis may result in a combination of real and complex conjugate modes. 

In that case, there is less redundancy among the $p$ modes, because each real mode is unique, but the number of extracted features needs to be exactly the same as it is when all the modes are complex conjugate pairs. To this end, we still extract $p'$ modes even though this requires removing some of the modes. More precisely, 
since the residue coefficients indicate the quantity of each mode present in each PMU data steam, we sort the modes based on their average residue across all the PMUs and we choose $p'=p/2$ modes with the largest average residues to be included in the vector of features. (The average residue corresponding to the $k\Th$ mode is $\frac{1}{m}\sum_{i=1}^{m}|R^{(i)}_k|$.) 
Based on our simulation results, we have found that $p=6$ and $p'=3$ is sufficient to ensure the accuracy and robustness of the estimated modes against noise (see Section \ref{sec:simul} for more details).

Moreover, since only a small portion of the PMUs ($m'< m$) capture the dynamic response of the system after an event, we only keep the residues of $m'$ PMUs with the largest magnitudes in the vector of features. Note that the $m'$ PMUs are not necessarily the same PMUs for different events.

Using the VPM channel measurements obtained from multiple PMUs, we define a row vector of features, $\mathcal{F}_{\text{VPM}}$, as follows:   
\begin{equation}\label{eq: VPM}
\begin{split}
 \mathcal{F}_{\text{VPM}} =     &[\{\omega_k:k=1,\ldots,p'\},\{\sigma_k:k=1,\ldots,p'\},\\
    &\{|R^{(i)}_k|:i=1,\ldots,m',k=1,\ldots,p'\}\\
    &\{\theta^{(i)}_k:i=1,\ldots,m',k=1,\ldots,p'\}]   
\end{split}
 \end{equation}
which consists of $p'$ angular frequencies, $p'$ damping factors and the corresponding magnitude and angle of the residues for each of the $m'$ PMUs (with the largest residue magnitudes) and $p'$ modes. 
To make a meaningful comparison of the features, it is important to sort them consistently. We sort the modes based on their average residue across all the $m'$ PMUs. In our notation in \eqref{eq: VPM}, $k=1$ represents the mode with the largest average residue and $k=p'$ represents the mode with the smallest average residue. Moreover, for a given mode $k$, the residues for different PMUs, $i=1,\ldots,m'$, are sorted in a descending order based on the magnitude of their residues, $|R^{(i)}_k|$ and we use the same order to sort the corresponding $\theta^{(i)}_k$. Note that, for each mode, we do not expect that the same PMU to always have the largest residue. Thus, the same PMU could be represented using a different index. 

In a similar manner, we obtain the set of features corresponding to other PMU channels, i.e., VPA, IPM, IPA, and F. 
Then each event $j$ can be described as a vector of features as   
\begin{equation}\label{eq: FEVPM}
\boldsymbol{\phi}_j = [\mathcal{F}_{\text{VPM}},\mathcal{F}_{\text{VPA}}, \mathcal{F}_{\text{IPM}},\mathcal{F}_{\text{IPA}}, \mathcal{F}_{\text{F}}]^T
\end{equation}
where each $\mathcal{F}_s$, $s \in \{\text{VPM,  VPA, IPM,  IPA, F}\}$ consists of the modal analysis results corresponding to the selected PMU channel. 
Hence, assuming $n_{ch}$ represents the number of channels at a PMU that are used for modal analysis, each event $j$ can be described as a set of $d$ features $\boldsymbol{\phi}_j = [\varphi_1, \dots, \varphi_d]^T\in \R^d$, where $d = 2 n_{ch} (p'+m'p')$. For instance, for $m'=25$,  $p'=3$, and using $n_{ch}=5$ channels, we obtain a total of $d = 780$ features. When the number of labeled events is small (e.g., 70 labeled events in our proprietary dataset) which is typically the case in practice, a 780-dimensional feature set can be extremely large. 

\subsection{Feature selection using filter methods}\label{sec: FS}

 Although there exist many different filter methods in the literature \cite{Filter-review}, in this paper, rather than measuring the interdependence among the features, we only focus on measuring the dependence between features and the target variables to rank the features and retain the top ranked ones.
 As the measure of dependence, various statistical tests, including one-way analysis of variance F-value test, sure independence screening, mutual  information,  Pearson correlation, and Kendall correlation have been used in literature \cite{zebari2020comprehensiveFS}. Given that we are focusing on a classification setting, we are interested in determining the correlation between numerical features and a categorical target variable. To this end, we use F-value test (F)\cite{sheikhan2013modular}, sure independence screening (S)\cite{fan2008sure}, and mutual information (M) \cite{ross2014MI_SKlearn} to quantify the correlation between features and the target variable. We use the off-the-shelf packages in Python to estimate the mutual information between discrete and continuous variables based on the nearest neighbor method (see \cite{ross2014MI_SKlearn} for more details).

As detailed in \ref{sec:Pre-Proc}, each event $j$ can be described as a set of $d$ features $\boldsymbol{\phi}_j = [\varphi_1, \dots, \varphi_d]^T\in \R^d$ and a label $\xi_j$ which describes the class of the event (i.e., line trips and generation loss events are labeled as 0 and 1, respectively). We define our dataset, $D = \{\boldsymbol{\phi}_j, \xi_j\}_{j=1}^{N_{\text{e}}}$ where $N_\text{e}$ is the total number of labeled events. We use the Z-score to normalize our dataset \cite{friedman2001elements}. Then we split the dataset into a training dataset with $N_\text{tr}$ samples and a test dataset with $N_\text{te}$ samples, denoted as $D_{\text{train}}$ and $D_{\text{test}}$, respectively. 
In a standard filter method, we compute the correlation of each feature $\varphi_i$, $i=1,\ldots,d$ and the target variable, $\boldsymbol{\Xi}=[\xi_1,\dots,\xi_j,\dots,\xi_{N_\text{tr}}]^T \in \{0,1\}^{N_\text{tr}}$ in the training dataset. Then we sort the features based on their correlation measure and then keep the $d'$ features with the highest correlation.

However, due to the small number of samples, we need a more robust way of choosing the features. Therefore we will rely on a well-known approach in machine learning, bootstrapping. Bootstrapping is a technique of sampling with replacement to create multiple datasets from the original dataset, thereby selecting the most informative features with some degree of statistical confidence. 
Note that the size of each bootstrapped dataset is the same as the original dataset.

\begin{figure}[t!]
\centering
\includegraphics[trim = {5.7cm 0 3cm 0},clip,scale=0.45]{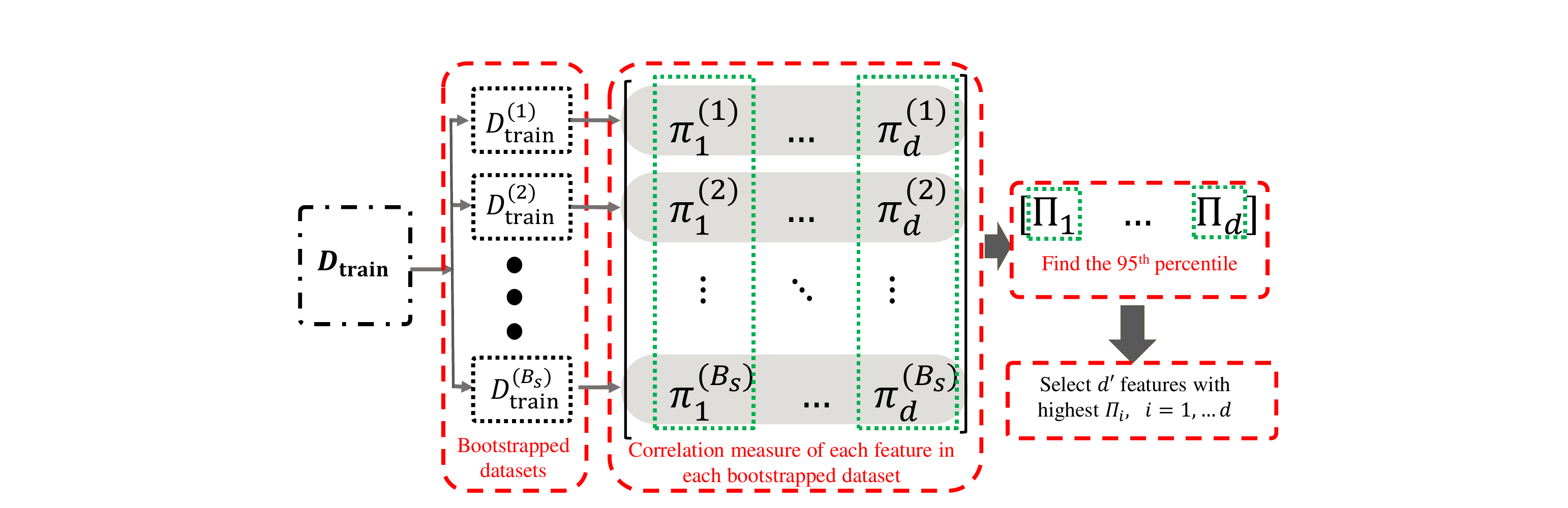}
\caption{Overview of the feature selection step} 
\label{fig:filter}
\end{figure}

\textbf{The proposed bootstrapping approach for feature selection:} The overview of feature selection step is shown in Fig. \ref{fig:filter}. The process begins by constructing $B_s$ bootstrapped datasets, denoted as $D^{(b)}_{\text{train}}$, $b=1,\ldots,B_s$, from the original training dataset, $D_{\text{train}}$. For each bootstrapped dataset, we randomly select $N_{\text{tr}}$ events from the training dataset. Each event in the original training dataset has an equal probability of being included in a bootstrapped dataset and can be included more than once or not at all. In other words, for each bootstrapped dataset, we sample from the original dataset with replacement. 
 We define, $\pi_{i}^{(b)}$ as the correlation measure of feature $\varphi_i$ and target variable $\boldsymbol{\Xi}$ over the $b\Th$ bootstrap samples. 
  In order to robustly find a subset features, we compute the $95$th percentile of the correlation measures of each feature over the $B_s$ bootstrapped datasets and select $d'$ features with the highest $95$th  percentiles. Using the selected $d'$ features, we obtain a reduced order training dataset, denoted as $D'_{\text{train}}$. 

We will also use bootstrapping for the classification (see Section \ref{sec: valid}).  We have done extensive experiments without bootstrapping which confirms the advantage of using it for both feature selection as well as the classification. In the interest of clarity, we did not include those results in this paper. 

\section{Event Classification}\label{sec: valid}
 
The final step in our proposed framework for event identification is to use the subset of features (as described in Section \ref{sec: FE}) to learn a classification model by finding decision boundaries between various event classes in the feature space.  With any ML model, there is a tradeoff inherent in the choice of complexity of the classification model. A simpler model may be more easily interpreted and is less likely to encounter overfitting problems whereas a more complex model may be more capable of uncovering subtle characteristics of the underlying phenomena and may thereby perform better. Therefore, to investigate the impact of the model complexity on the accuracy of event classification, two well-known classifiers, namely, LR and SVM with RBF kernels are used to identify the two classes of events in our dataset (we refer readers to \cite{friedman2001elements} for details of the two classification models). The LR is a relatively simple model compared to the SVM with RBF kernels.

In order to validate the performance of each classification model, we split the dataset into a training and a test datasets. All the filter methods are implemented on the training dataset to find the most relevant and informative subset of features and obtain reduced order training and test datasets, denoted as $D'_{\text{train}}$ and $D'_{\text{test}}$, respectively. Due to the limited number of labeled generation loss and line trip events, we again use the bootstrap technique as a tool for assessing statistical accuracy. Using bootstrap sampling helps to address the problem of limited training samples and therefore justifies using the test data for validation of specific parameters, namely, the number of features to pick and the choice of the classification model. 

\begin{figure}[t!]
\centering
\includegraphics[trim = {2.5cm 0 2cm 0},clip,scale=0.3]{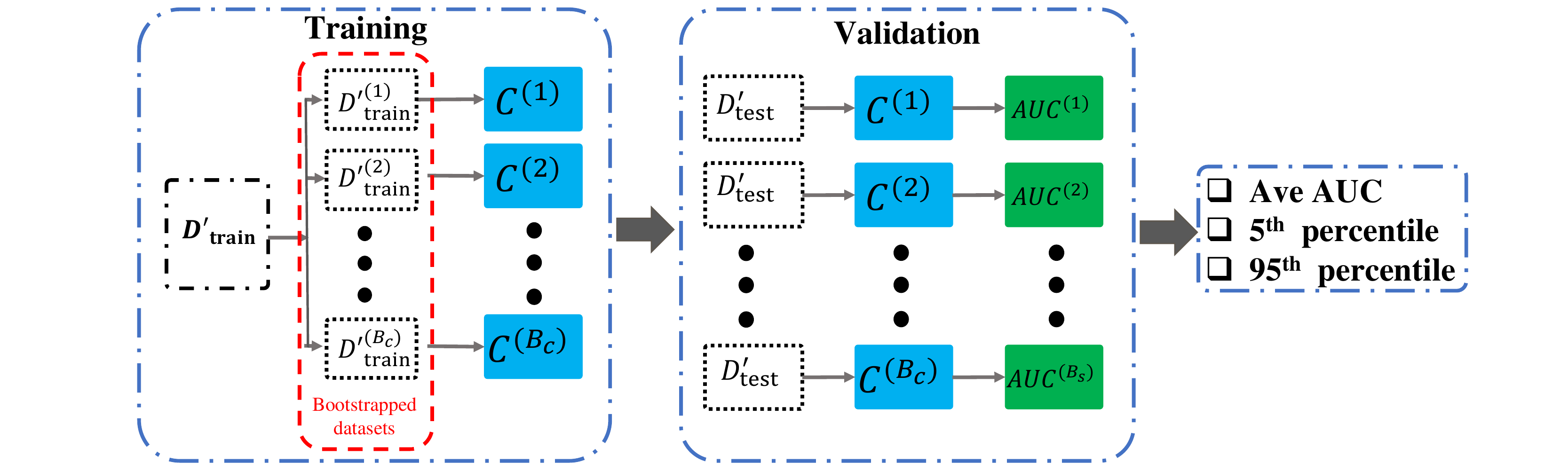}
\caption{Overview of the model validation ($D'_{\text{train}}$, and $D'_{\text{test}}$ are reduced order training and test data, respectively. $C^{(i)}$: learned model from the $i\Th$ bootstrapped reduced order training data.)} 
\label{fig:valid}
\end{figure}

Using the reduced order training dataset, $D'_{\text{train}}$, we generate $B_c$ reduced order bootstrapped datasets, denoted as $D'^{(b)}_{\text{train}}$, $b=1,\ldots,B_c$, to learn a classification model, $C^{(b)}$, and classify the events in the $D'_{\text{test}}$. 
To evaluate the performance of a chosen classifier (for example, LR), we use the area under curve (AUC) of the receiver operator characteristic (ROC), which characterizes the accuracy of the classification for various discrimination thresholds \cite{zebari2020comprehensiveFS}. (The discrimination threshold determines the probability at which the positive class is chosen over the negative class.) The ROC plot shows the relation between the true positive rate and the false positive rate at various threshold settings.   The ROC AUC value is bounded between 0 and 1. The closer AUC to 1, the classifier has a better ability to classify the events. To quantify the accuracy of the learned classifier on the test dataset, we compute the average AUC, and the corresponding 5th and 95th percentiles of the AUC values over all the bootstrapped datasets. The aforementioned steps are summarized in Fig. \ref{fig:valid}.

\section{Simulation results}\label{sec:simul}

 In order to evaluate the performance of the proposed framework for event identification, two different datasets are considered in this study. 
 The first one is obtained from the dynamic simulation of line trip and generation loss events in the Texas 2000-bus synthetic grid \cite{Texas2000} using the power system simulator for
engineering (PSS$^{\circledR}$E). 
 The second dataset is a proprietary dataset with labeled generation loss and line trip events obtained from a large utility in the USA involving measurements from nearly 500 PMUs. 
 
 In the remainder of the section, we present our results for each dataset including: (i) the sufficient number of distinct modes, $p'$, using the measurements obtained from different PMU channels, (ii) the reconstruction error of the PMU measurements using modal information obtained from MSMPM, , (iii) the performance of LR and SVM in identifying the events using the subset of features (as explained in Sections \ref{sec: Math} and \ref{sec: FE}),  and (iv) we compare the performance of our proposed approach with a well established event identification method based on the work of Li et al. \cite{li2018real}.


\subsection{Case 1: Event classification with the synthetic dataset}

In order to generate synthetic PMU data with labeled events, we use the PSS$^{\circledR}$E dynamic data of the the Texas 2000-bus synthetic grid \cite{Texas2000Data}.
We allow the system to be in the normal operation condition for $1$ second. Then, we apply a line trip or generation loss at time $t=1$ and run the dynamic simulation to $t=20$ seconds. 
 The simulation time step for dynamic simulations is set to $0.0083$ secs. In order to collect data at a rate of 30 sample/sec (PMU sampling rate), we record the measurements at each $0.033 / 0.0083 \approx 4$ time steps. We assume that 95 of the 500 kV buses (which are chosen randomly) across the grid are equipped with PMU devices. 
We generate a total number of 800 events including 400 generation loss and 400 line trip events. For each event class, 200 events are simulated under the normal loading and 200 with $80\%$ of normal loading. 
Since PSS$^{\circledR}$E does not have any channel to directly measure the branches currents, only VPM, VPA, and F channels are used for extracting the features from the PMU measurement. To capture the dynamic response of the system, we use $N=300$ samples after the exact start of an event. 

To evaluate the performance of the classification models, we split our synthetic data into training and test datasets with $600$ and $200$ samples, respectively. The training dataset is used for feature engineering and learning the models and the test dataset is only used for evaluation and comparison of the models.

\begin{figure}[t!]
\centering
\includegraphics[scale=0.4]{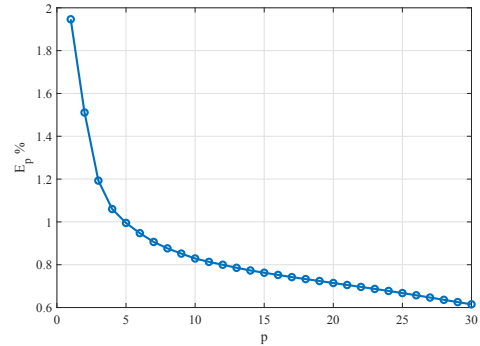}
\caption{Rank $p$ approximation error of the matrix $\mathbf{H}$ (which is obtained from VPM measurements from $95$ PMUs after a line trip event) for different values of $p$.}
\label{fig:Hankel_Syn}
\end{figure}

Using the VPM measurements obtained from $95$ PMUs after a line trip event, we construct the matrix $\mathbf{H}$ based on \eqref{eq:H_i}. 
In Fig. \ref{fig:Hankel_Syn}, we illustrate the rank $p$ approximation error of the matrix $\mathbf{H}$ that is given by \eqref{eq:E_p}. The matrix $\mathbf{H}$ is constructed over a block of $N=300$ samples after the exact start time of the event with the pencil parameter of $L=150$. Observe that if one chooses a threshold of $1\%$ for the approximation error, then we only require $p=6$ largest singular values; this is the case for all the events in our synthetic dataset.

\begin{figure}[t!]
\centering
\hspace*{-0.5cm}  
\includegraphics[scale=0.4]{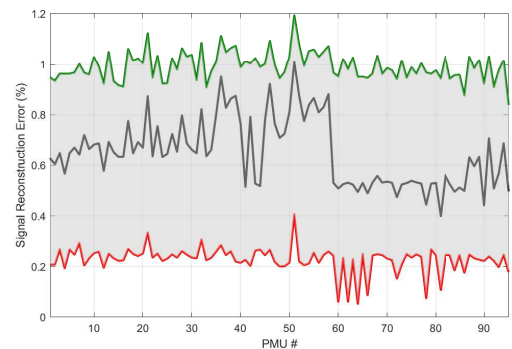}
\caption{Envelope of the reconstruction error of all the PMU measurement streams that are obtained from VPM channel after 800 events in our dataset. Red, gray, and green lines represent the minimum, average and maximum reconstruction error, respectively. }
\label{fig:E_i_envelop}
\end{figure}

 Fig. \ref{fig:E_i_envelop} illustrates the envelope of the reconstruction error of all the PMU measurement streams (that are obtained from VPM channel) in the synthetic dataset.  The average reconstruction error of the PMUs over all the events in our dataset is less than $1\%$. As detailed in the Section \ref{sec:Hankel}, this implies that using $p=6$ modes is sufficient for capturing the underlying dynamics of the system after an event.  
  



As discussed in Section \ref{sec: FE}, to remove the redundant information present in the complex conjugate modes, we use $p'=3$ distinct modes in the vector of features for each event. Furthermore, to determine the parameter $m'$, 
we use the normalized residue for each PMU with respect to the one with the largest magnitude and pick the smallest number of PMUs for which more than $95\%$ of the PMUs are less than a certain threshold. Based on this approach, we choose $m'=20$ PMUs to capture the most significant residues in our synthetic dataset. 
Therefore, considering $p'=3$, $m'=20$, and $n_{ch}=3$, each event in the synthetic dataset is characterized using $d = 378$ features. 
Then, we generate $B_s=200$ bootstrapped datasets from the original training dataset to retain the features with the highest correlation. 



\begin{figure*}[h!]
\centering
\begin{subfigure}{0.5\textwidth}
  \centering
  \includegraphics[trim={0 0 0 500},clip,scale=0.52]{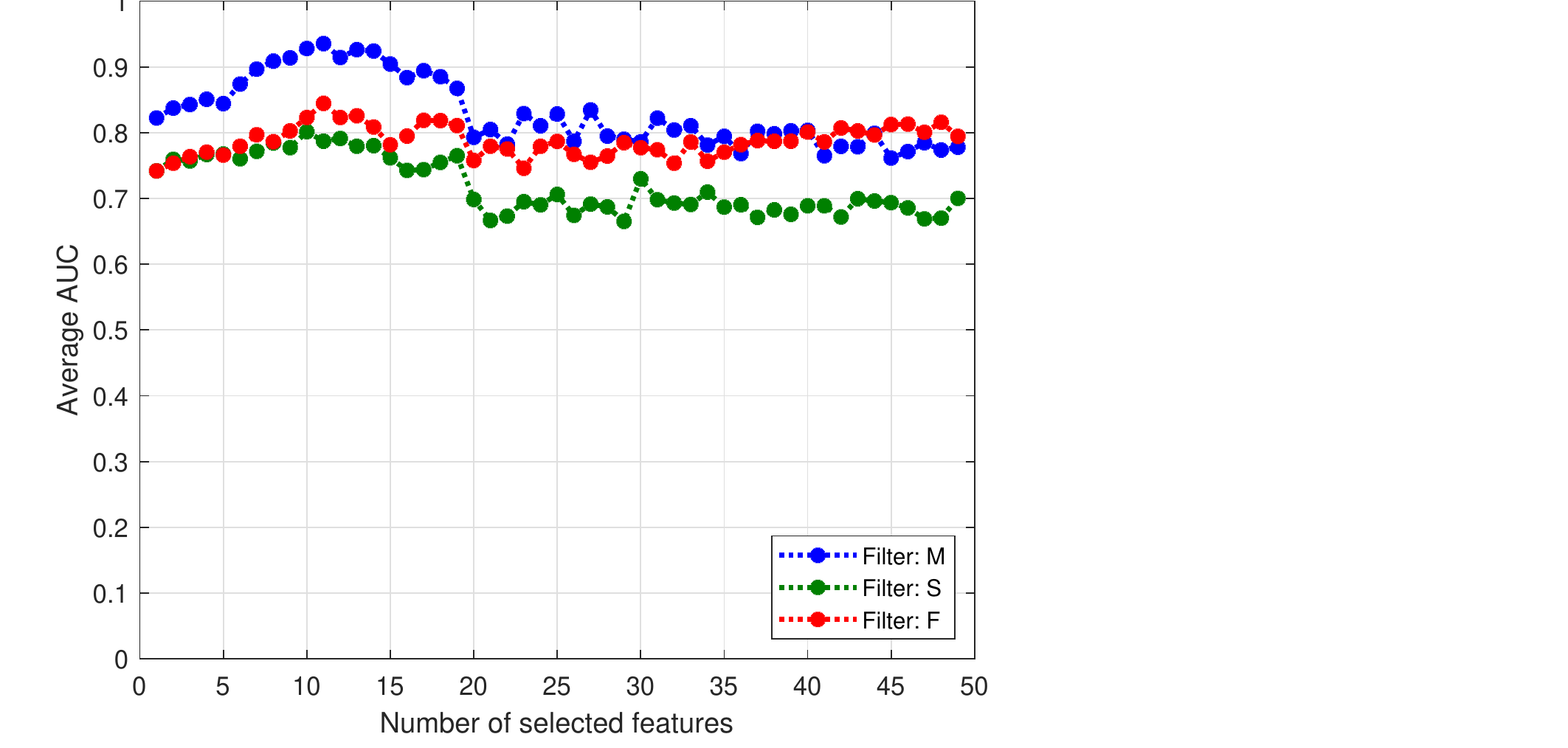}
  \caption{}
  \label{fig:sub1}
\end{subfigure}%
\begin{subfigure}{0.5\textwidth}
  \centering
  \includegraphics[trim={0 0 0 500},clip,scale=0.52]{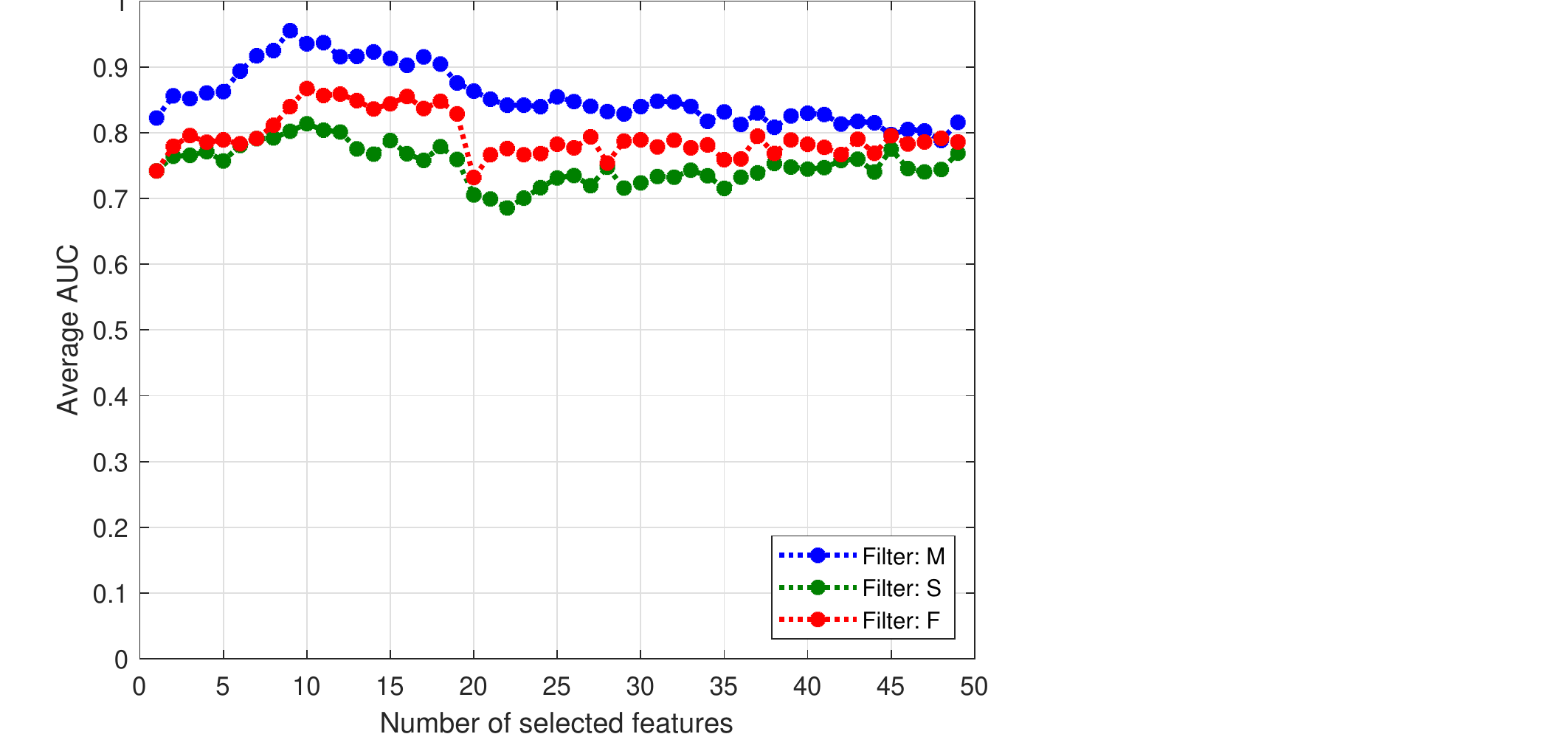}
  \caption{}
  \label{fig:sub2}
\end{subfigure}
\caption{Performance of the classification models (a) LR, and (b) SVM in terms of average AUC over $B_c=200$ bootstrapped datasets with respect to the number of selected features in the synthetic dataset. 
}
\label{fig:synthetic}
\label{fig:LR_dp}
\end{figure*}

\begin{figure*}[h!]
\centering
\begin{subfigure}{0.5\textwidth}
  \centering
  \includegraphics[scale=0.52]{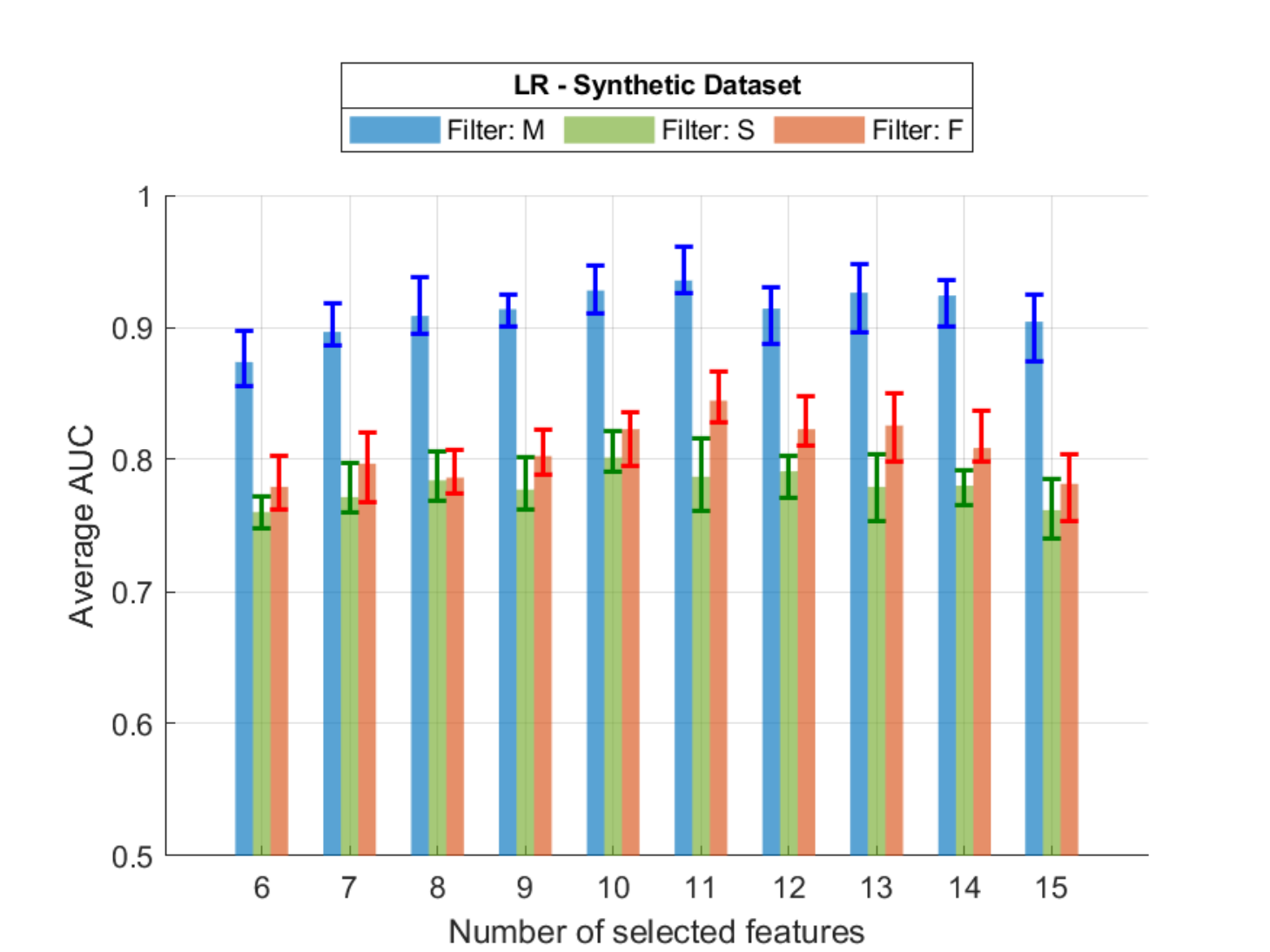}
  \caption{}
  \label{fig:sub1}
\end{subfigure}%
\begin{subfigure}{0.5\textwidth}
  \centering
  \includegraphics[scale=0.52]{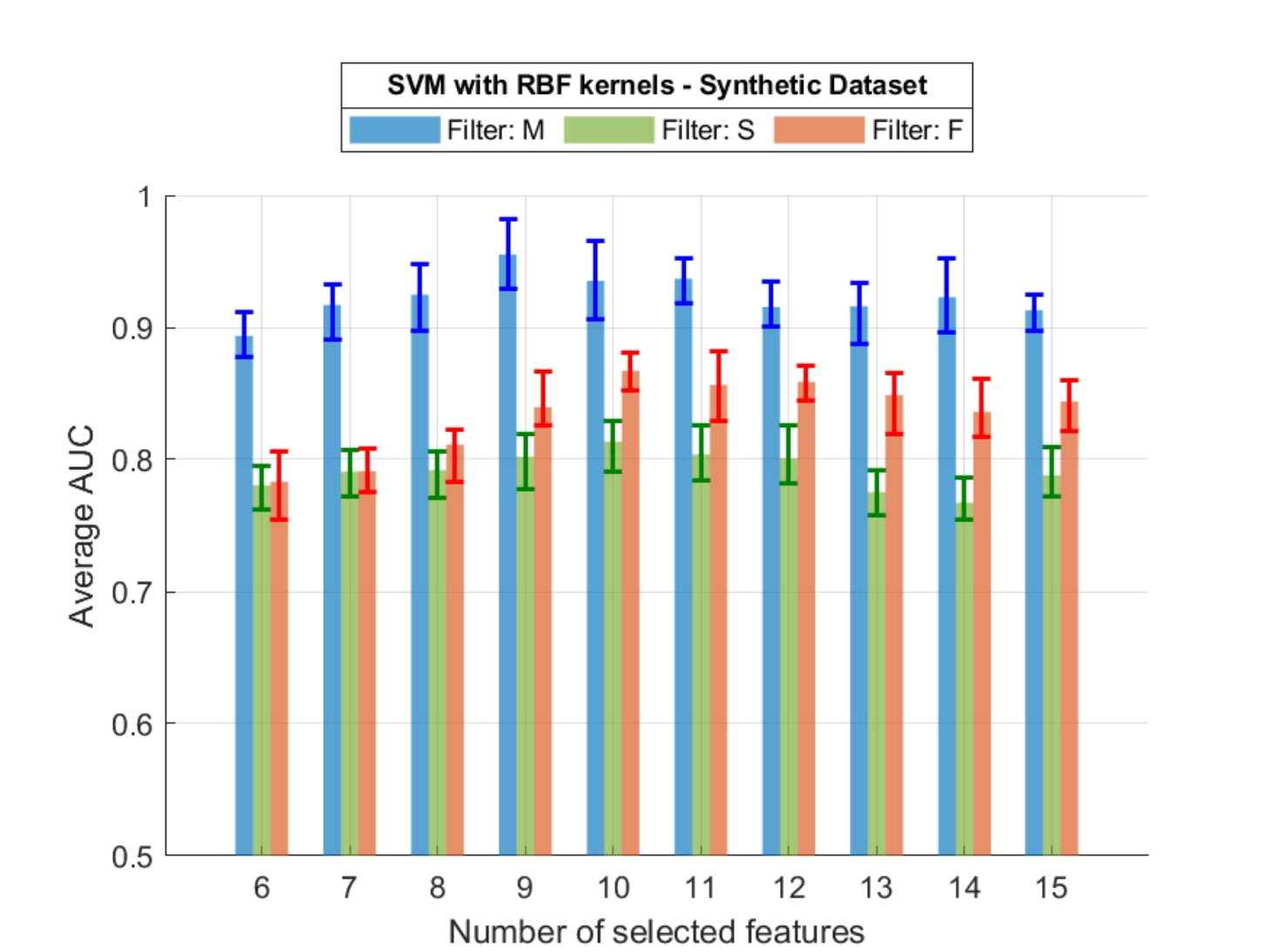}
  \caption{}
  \label{fig:sub2}
\end{subfigure}
\caption{Performance of the classification models (a) LR, and (b) SVM in terms of the average AUC with respect to the number of selected features in the synthetic dataset. The error bars represent the 5th and 95th percentiles of the AUC scores.}
\label{fig:synthetic}
\end{figure*}

Figure \ref{fig:LR_dp} shows the performance of the classification models, namely, (a) LR, and (b) SVM in terms of average AUC over $B_c=200$ bootstrapped datasets with respect to the number of selected features.   The selected features are the ones with the highest 95th percentiles obtained from various  correlation measures (i.e., F, S, and M as detailed in Section \ref{sec: FS}). To further elaborate the performance of each classifier, using a subset of $6$ to $15$ features obtained from various correlation measures, the average AUC score as well its corresponding 5th and 95th confidence intervals are shown in Fig. \ref{fig:synthetic}.

Based on the simulation results, using the mutual information as the correlation measure to select a subset of features will result in a better performance of both classifiers. This is due to the fact that F-value and sure independence screening only consider the linear dependence of the features with the target variable whereas mutual information can also capture non-linear dependencies. The selected features include the angular frequency and first few residue magnitudes corresponding to the first mode of the VPM, VPA, and F measurement channels.  
Furthermore, it is clear that SVM with RBF kernel has a slightly better performance than LR in identifying the two classes of the events in our synthetic dataset.  It is also clear that using a subset of about $10$ features obtained from mutual information will result in the best performance of both classifiers. The error bars represent the 5th and 95th percentile of the AUC scores over $B_c$ bootstrapped datasets and are an indication of the robustness of each learned classifier.

\begin{figure*}[t!]
\centering
\begin{subfigure}{0.5\textwidth}
  \centering
  \includegraphics[trim={0 0 0 480},scale=0.52]{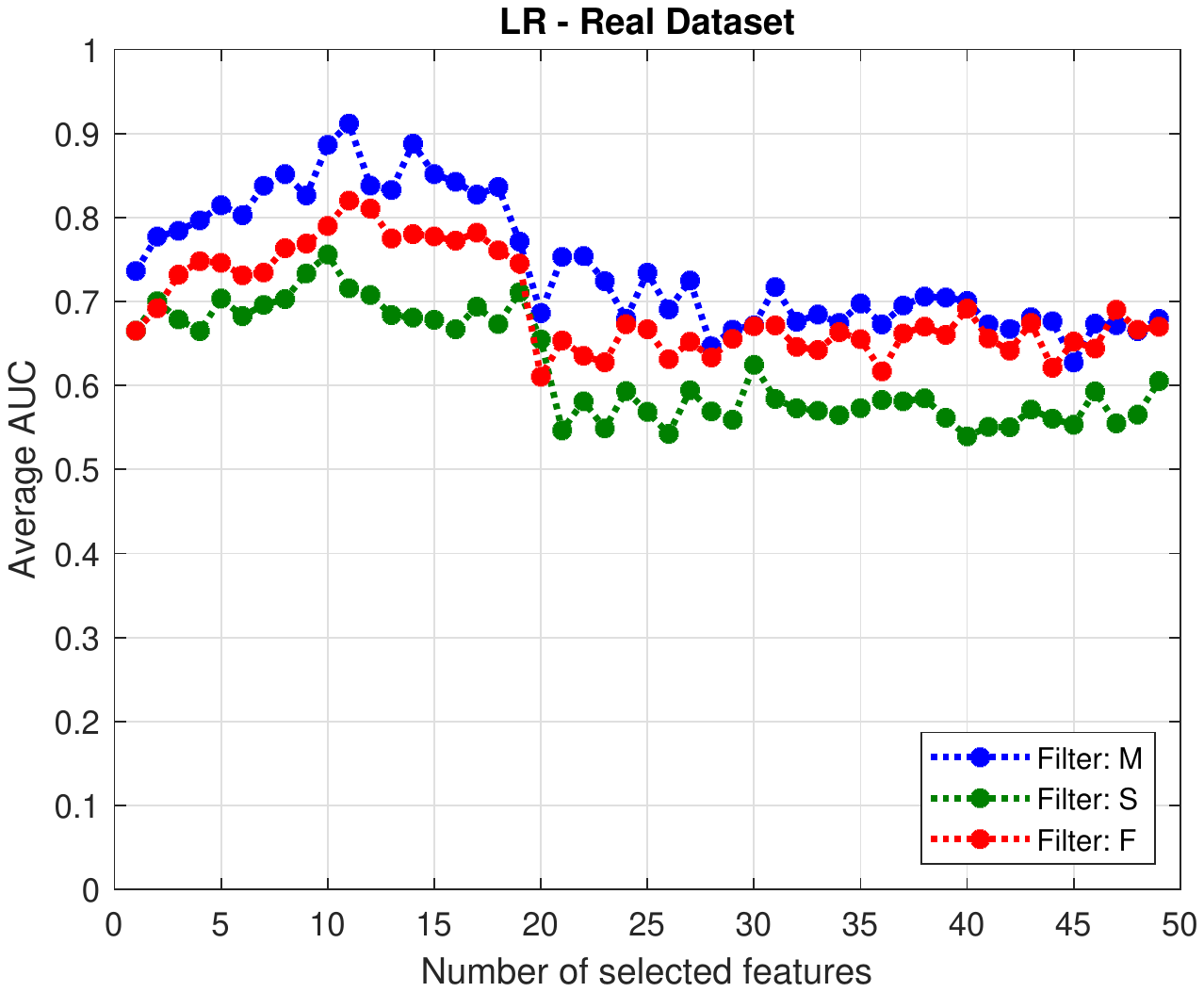}
  \caption{}
  \label{fig:sub1}
\end{subfigure}%
\begin{subfigure}{0.5\textwidth}
  \centering
  \includegraphics[scale=0.52]{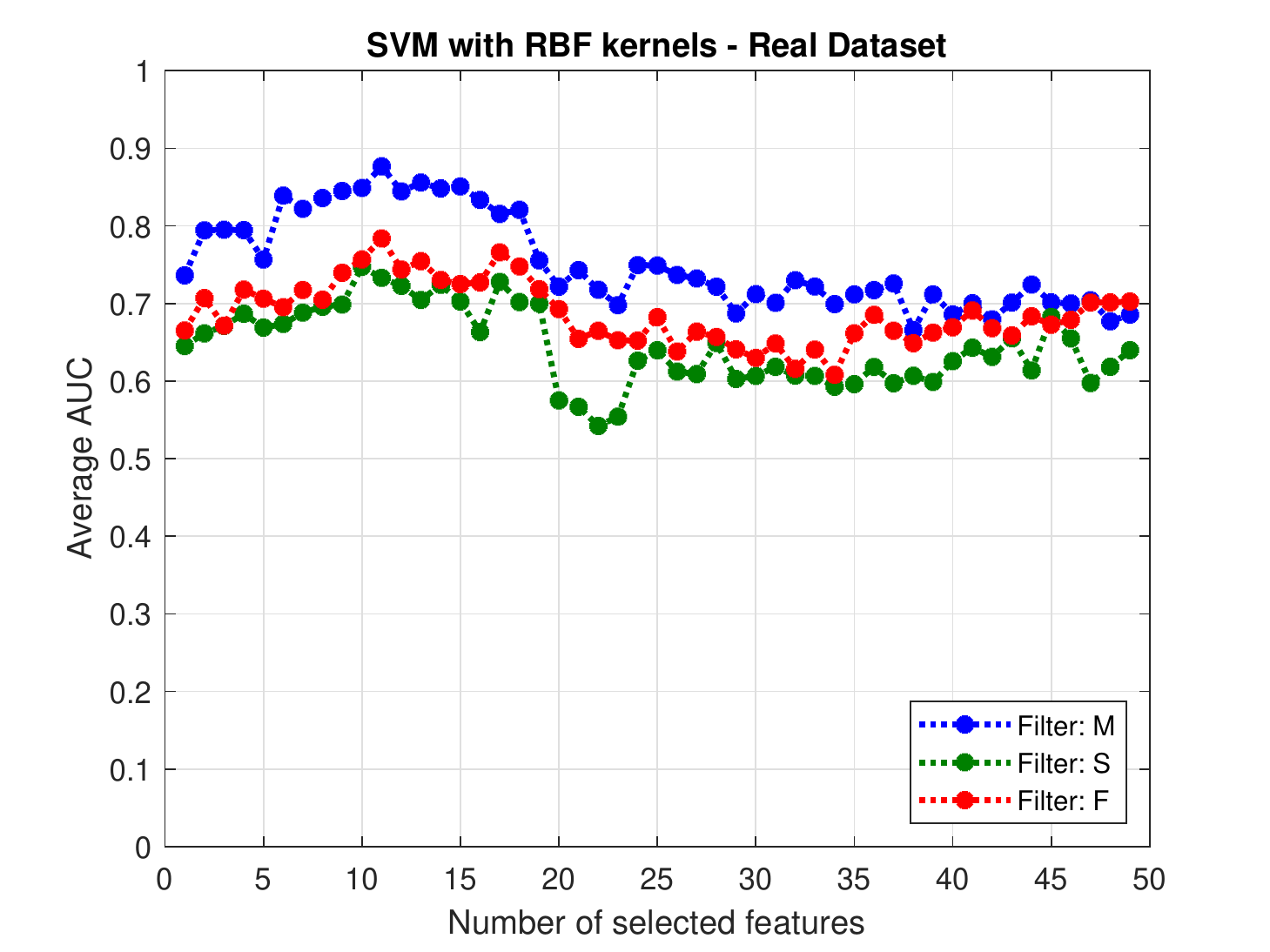}
  \caption{}
  \label{fig:sub2}
\end{subfigure}
\caption{Performance of the classification models (a) LR, and (b) SVM in terms of average AUC over $B_s=200$ bootstrapped datasets with respect to the number of selected features in the real dataset. 
}
\label{fig:Real_AUC}
\end{figure*}

\begin{figure*}[t!]
\centering
\begin{subfigure}{0.5\textwidth}
  \centering
  \includegraphics[scale=0.52]{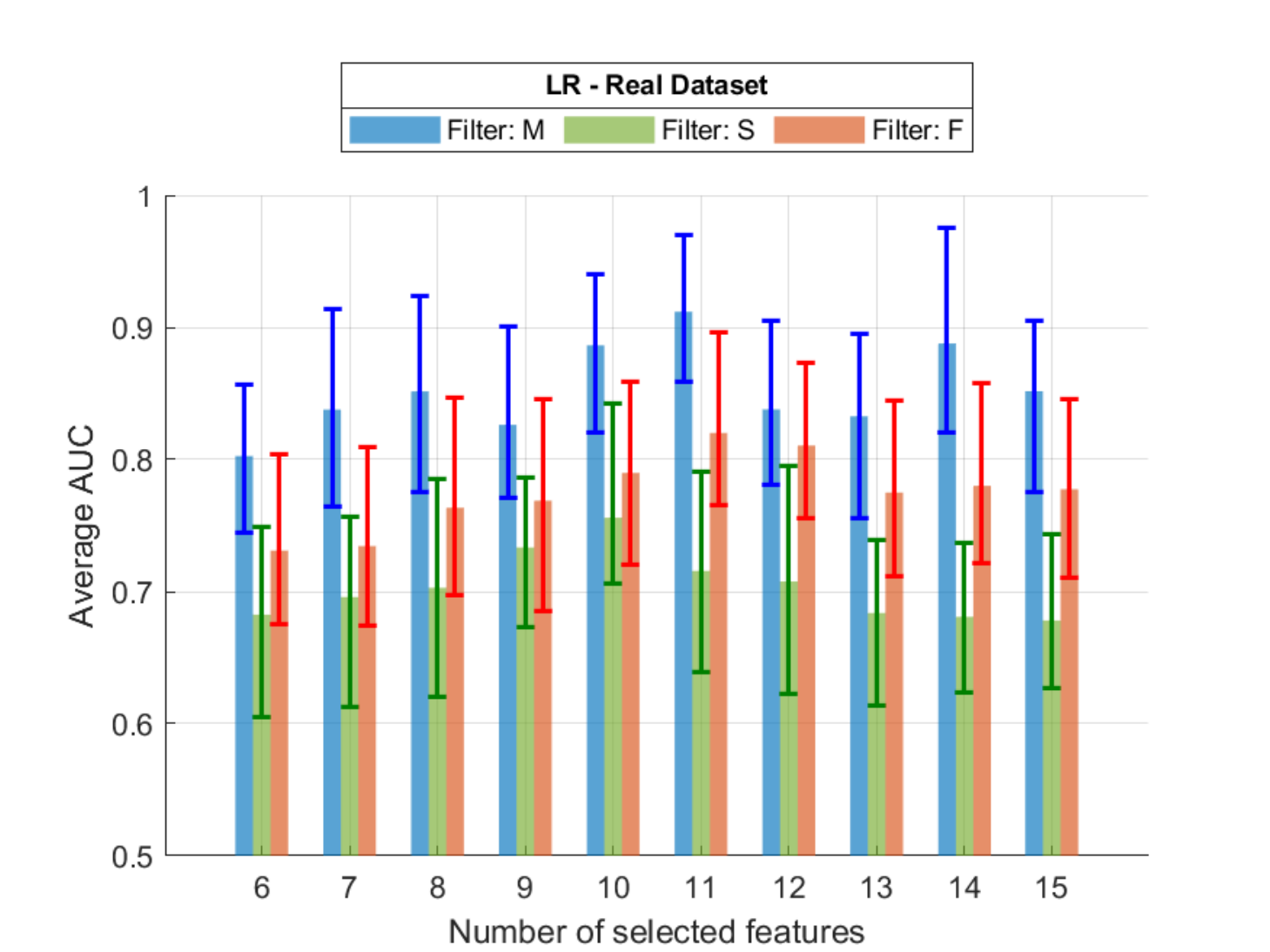}
  \caption{}
  \label{fig:sub1}
\end{subfigure}%
\begin{subfigure}{0.5\textwidth}
  \centering
  \includegraphics[scale=0.52]{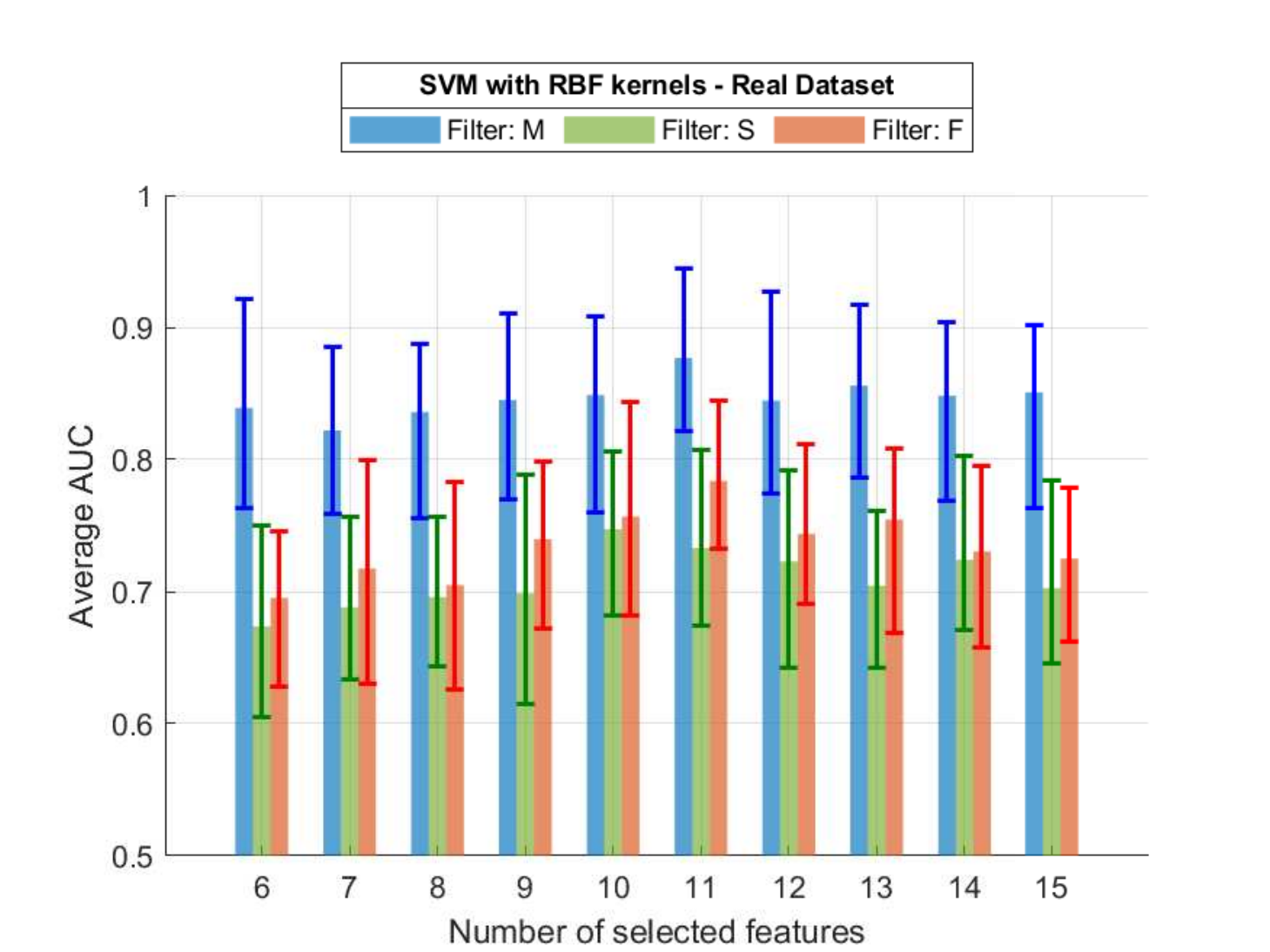}
  \caption{}
  \label{fig:sub2}
\end{subfigure}
\caption{Performance of the classification models (a) LR, and (b) SVM in terms of the average AUC with respect to the number of selected features in the real dataset. The error bars represent the 5th and 95th percentiles of the AUC scores.}
\label{fig:Real_AUC_CI}
\end{figure*}

\subsection{Case 2: Event classification with a proprietary datset}

\begin{figure}[h!]
\centering
\hspace*{-0.5cm}  
\includegraphics[scale=0.45]{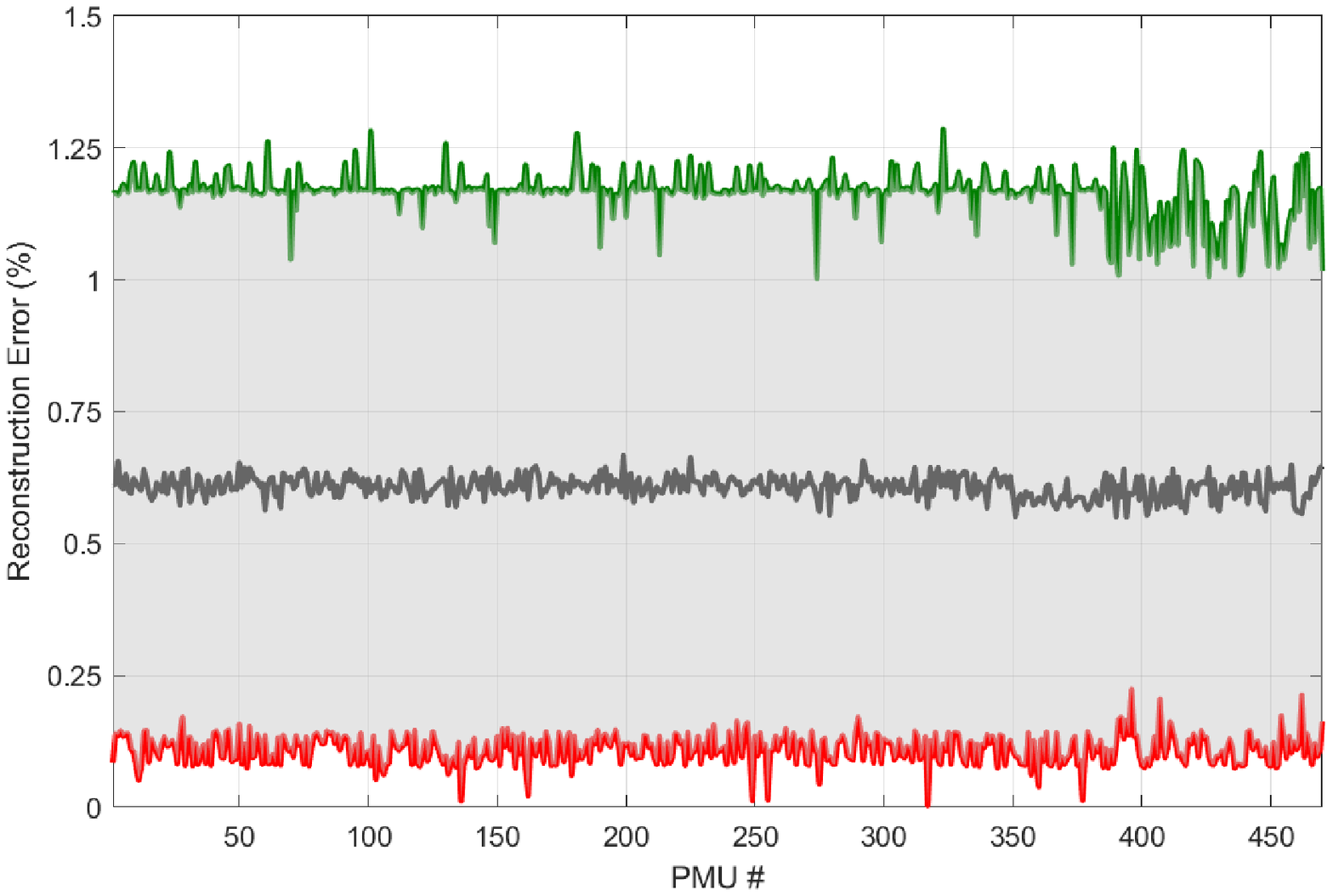}
\caption{Envelope of the reconstruction error of all the PMU measurement streams that are obtained from VPM channel after 70 events in our real datasets. Red, gray, and green lines represent the minimum, average and maximum reconstruction error, respectively.}
\label{fig:E_i_envelop_r}
\end{figure}

To further investigate the performance of our proposed framework, we use a proprietary PMU data obtained from a large utility in the USA involving measurements from nearly 500 PMUs. 
A total of 70 labeled events including 23 generation loss and 47 line trip events are used in this study. To characterize the dynamic response of the system after an event, VPM, VPA, IPM, IPA, and F measurement channels from multiple PMUs over a block of $N=300$ samples (after the exact start time of the event) are used for extracting the features as discussed in Section \ref{sec: Math}. The envelope of the reconstruction error of all the PMU measurement streams (that are obtained from VPM channel) in the synthetic dataset.
Fig. \ref{fig:E_i_envelop_r} illustrates that using $p=6$ modes, the average reconstruction error of the PMUs over all the events in our real dataset is less than $1\%$. 
Using the same approach that is used in case 1, the parameters $p'=3$ and $m'=25$ are used to construct the vector of features for each event, thereby obtaining a total number of $d=780$ features.

A total number of 56 events are included in the training dataset. The same number of $B_s=200$ bootstrapped datasets are used for feature selection and the final evaluation of the models. The performance of each classifier in terms of the average AUC scores are shown in Fig. \ref{fig:Real_AUC}. Further, the 5th and 95th percentiles of the AUC scores over $B_s=200$ bootstrapped datasets are shown in Fig. \ref{fig:Real_AUC_CI}. 

The best performance of the both classifiers are obtained using a subset of $11$ features that are selected based on the mutual information. An interesting observation is that in both case studies, the angular frequency and first few residue magnitudes corresponding to the first mode of VPM, VPA and F measurement channels are included in the subset of the selected features obtained from mutual information.

Compared to the synthetic dataset, the performance of the classification models in the real dataset have lower accuracies with wider confidence intervals. 
Possible reasons for this include (i) the limited number of events ($70$ labeled events), and (ii) variable system operating conditions as the data was collected over $3$ years. 
Furthermore, in contrast to our simulation results for the synthetic dataset, the learned LR model demonstrates a slightly better performance compared to the learned SVM with RBF kernels model. This is most likely because SVM significantly increases the model complexity and given the small number of samples, is overfitting the training dataset and thus, will not perform as well on the test data \cite{friedman2001elements}. 

In terms of the time needed to identify an event using our proposed approach, once an event is detected, we only need to extract the features corresponding to various PMU channels using the multi-signal MPM and feed the selected subset of features into our learned model. Based on our simulations, this procedure takes about 1.1 second. The simulations are conducted on a computer with 8 GB RAM and Intel Core i5 processor with 1.6 GHz CPU.
\begin{table*}[h!]
\caption{\label{tab:comparison} Comparison of the confusion matrices for the real dataset with 23 generation loss (GL) and 47 line trip (LT) events and the synthetic dataset with 400 GL and 400 LT events.}
\centering
\begin{tabular}{|c|cccc|cccc|c|}
\hline
\multirow{3}{*}{} & \multicolumn{4}{c|}{\textbf{\begin{tabular}[c]{@{}c@{}}Based on low-dimensional \\ subspace \cite{li2018real}\end{tabular}}}                                                                                                                                 & \multicolumn{4}{c|}{\textbf{Our Proposed   Method}}                                                                                                                                                                                                                          & \multirow{3}{*}{\textbf{Dataset}} \\ \cline{2-9}
                  & \multicolumn{2}{c|}{Parameters in \cite{li2018real}}                                                                            & \multicolumn{2}{l|}{Tuned Parameters}                                                                                      & \multicolumn{2}{c|}{VPm   channel}                                                                                                             & \multicolumn{2}{c|}{All channels}                                                                                           &                                   \\ \cline{2-9}
                  & \multicolumn{1}{c|}{LT}                                      & \multicolumn{1}{c|}{GL}                                                           & \multicolumn{1}{c|}{LT}                                      & GL                                                          & \multicolumn{1}{c|}{LT}                                     & \multicolumn{1}{c|}{GL}                                                          & \multicolumn{1}{c|}{LT}                                      & GL                                                           &                                   \\ \hline
LT                & \begin{tabular}[c]{@{}c@{}}\textbf{63.83 \%}\\ (30/47)\end{tabular}   & \multicolumn{1}{c|}{\begin{tabular}[c]{@{}c@{}}\textbf{36.17 \%}\\ (17/47)\end{tabular}}   & \begin{tabular}[c]{@{}c@{}}\textbf{72.35 \%}\\ (34/47)\end{tabular}   & \begin{tabular}[c]{@{}c@{}}\textbf{27.65 \%}\\ (13/47)\end{tabular}  & \begin{tabular}[c]{@{}c@{}}\textbf{78.72 \%}\\ (37/47)\end{tabular}  & \multicolumn{1}{c|}{\begin{tabular}[c]{@{}c@{}}\textbf{21.27\%}\\ (10/47)\end{tabular}}   & \begin{tabular}[c]{@{}c@{}}\textbf{80.85 \%}\\ (38/47)\end{tabular}   & \begin{tabular}[c]{@{}c@{}}\textbf{19.15\%}\\ (9/47)\end{tabular}     & \multirow{2}{*}{Real}             \\ \cline{1-1}
GL                & \begin{tabular}[c]{@{}c@{}}\textbf{39.13 \%}\\ (9/23)\end{tabular}    & \multicolumn{1}{c|}{\begin{tabular}[c]{@{}c@{}}\textbf{60.87\%}\\ (14/23)\end{tabular}}    & \begin{tabular}[c]{@{}c@{}}\textbf{21.73\%}\\ (5/23)\end{tabular}     & \begin{tabular}[c]{@{}c@{}}\textbf{78.27\%}\\ (18/23)\end{tabular}   & \begin{tabular}[c]{@{}c@{}}\textbf{26.01 \%}\\ (6/23)\end{tabular}   & \multicolumn{1}{c|}{\begin{tabular}[c]{@{}c@{}}\textbf{73.99 \%}\\ (17/23)\end{tabular}}  & \begin{tabular}[c]{@{}c@{}}\textbf{21.73\%}\\ (5/23)\end{tabular}     & \begin{tabular}[c]{@{}c@{}}\textbf{78.27\%}\\ (18/23)\end{tabular}    &                                   \\ \hline
LT                & \begin{tabular}[c]{@{}c@{}}\textbf{80.25 \%}\\ (321/400)\end{tabular} & \multicolumn{1}{c|}{\begin{tabular}[c]{@{}c@{}}\textbf{19.75 \%}\\ (79/400)\end{tabular}}  & \begin{tabular}[c]{@{}c@{}}\textbf{89.25 \%}\\ (357/400)\end{tabular} & \begin{tabular}[c]{@{}c@{}}\textbf{10.75 \%}\\ (43/400)\end{tabular} & \begin{tabular}[c]{@{}c@{}}\textbf{88.5 \%}\\ (354/400)\end{tabular} & \multicolumn{1}{c|}{\begin{tabular}[c]{@{}c@{}}\textbf{11.5 \%}\\ (46/400)\end{tabular}}  & \begin{tabular}[c]{@{}c@{}}\textbf{90.25 \%}\\ (361/400)\end{tabular} & \begin{tabular}[c]{@{}c@{}}\textbf{9.75 \%}\\ (39/400)\end{tabular}   & \multirow{2}{*}{Synthetic}        \\ \cline{1-1}
GL                & \begin{tabular}[c]{@{}c@{}}\textbf{20.75 \%}\\ (83/400)\end{tabular}  & \multicolumn{1}{c|}{\begin{tabular}[c]{@{}c@{}}\textbf{79.25 \%}\\ (317/400)\end{tabular}} & \begin{tabular}[c]{@{}c@{}}\textbf{15.5 \%}\\ (62/400)\end{tabular}   & \begin{tabular}[c]{@{}c@{}}\textbf{84.5 \%}\\ (338/400)\end{tabular} & \begin{tabular}[c]{@{}c@{}}\textbf{16.5 \%}\\ (66/400)\end{tabular}  & \multicolumn{1}{c|}{\begin{tabular}[c]{@{}c@{}}\textbf{83.5 \%}\\ (334/400)\end{tabular}} & \begin{tabular}[c]{@{}c@{}}\textbf{12.75 \%}\\ (51/400)\end{tabular}  & \begin{tabular}[c]{@{}c@{}}\textbf{87.25 \%}\\ (349/400)\end{tabular} &                                   \\ \hline
\end{tabular}
\end{table*}

\subsection{Comparison with prior work based on low-dimensional subspace of PMU data \cite{li2018real}}
To highlight the value of using multiple PMU channels and the combination of physics-based and data-driven methods in our proposed approach, we make a comparison with a well-established event identification method. This method was introduced in \cite{li2018real} and is based on  low-dimensional subspace characterization. In the following, we briefly explain the proposed approach in \cite{li2018real}.

Let $\hat{\mathbf{Y}} \in \R^{m\times N}$ be the data matrix which contains the PMU measurements from $m$ PMUs over a block of $N$ samples. 
The main idea in \cite{li2018real} is that events can be identified by the low-dimensional row subspace spanned by the dominant singular vectors of $\hat{\mathbf{Y}}$, denoted as  $V_r$, which consists of the right singular vectors associated with the $r$ largest singular values. For the task of event identification, a dictionary is constructed consisting of these $V_r$ matrices from labeled event data. The type of an event in the test dataset can be identified through comparing $V_r$ with the constructed dictionary. More specifically, the subspace angle between the span of their dominant right singular vectors is used to quantify this similarity.

Note that the choice of the parameters $r$ (number of dominant singular values) and $N$ (number of samples) in \cite{li2018real} is driven based on their dataset. In our implementation, in addition to using the original parameters in \cite{li2018real}, we also tuned the parameters to improve the performance in our datasets. Based on our simulations, we choose $r=5$ and $N=300$. 
Furthermore, the proposed approach in \cite{li2018real} uses a single PMU channel. Hence, we only use the VPm measurements to construct the PMU data matrix.

For the sake of comparison, we consider two scenarios in our proposed approach: in the first scenario we only use VPm channel measurements and in the second one, we use all the PMU channels to construct the vector of features.

To make a fair comparison, we consider a simple K-fold cross validation technique to compare the performance of each method in both real and synthetic datasets. We split each dataset into 5 folds and use 4 folds as the training dataset and the remaining fold as the test dataset. Hence, we obtain 5 different combination of training and test datasets and for each combination, we calculate a confusion matrix by combining all the test data (see Table \ref{tab:comparison}) to evaluate the performance of the methods. Note that as discussed above, we use the LR and SVM classifiers to identify the events in the real and synthetic dataset, respectively. 
Our simulations indicate that using a single PMU channel and tuning the parameters in \cite{li2018real}, the two methods have comparable performances. However, due to the fact that different PMU channels are able to capture the dynamic response of the system after an event, using all the PMU channels in our proposed approach results in a slightly better performance in identifying the events in both real and synthetic datasets.

\section{Concluding Remarks}\label{sec: concl}

We have proposed a novel machine learning framework for event identification based on extracted features obtained from mode decomposition of PMU measurements. Considering the high-dimensionality of the extracted features, we have considered different data-driven filter methods to choose a subset of features.  We have investigated the performance of the two classification models (LR and SVM) in identifying the generation loss and line trip events for both synthetic and a proprietary real datasets. It is worth noting that the reason for choosing only two types of events was the limited number of labeled events in our proprietary dataset. However, if the data for more than two types of event is available, the proposed approach can be easily extended to a multi-class classification problem by splitting it into a multiple binary classification problem.

Our simulation results indicate that using mutual information for feature selection results in better performance of the classifiers compared to the other filter methods that we tested, in both real and synthetic datasets. This is due to the fact that mutual information can capture the nonlinear dependencies between the features and the target variable. Our analysis also illustrates that bootstrapping can overcome the limitation of the small number of labeled events. However, when labeled data are limited, a less complex model such as LR can assure better accuracy than more complex models such as SVM. We have also shown that a relatively small number (10--15) of features is typically enough to achieve a good classification performance. 
While filter methods provide a streamlined way to identify the key features, in the future, we will explore alternative feature selection techniques such as least absolute shrinkage and selection operator (LASSO). The proprietary dataset used in this study suggests that, in practice, a very small number of events are labeled when compared to the total number of events. A possible solution to overcome this limitation is to incorporate the real labeled events in combination with the synthetic labeled events obtained by running PSS$^{\circledR}$E simulations on the same system model. Building on this, an interesting potential application of the proposed methodology is to the semi-supervised setting wherein labeled events are combined with unlabeled data streams in which an event has been detected, but for which the class of event is unknown, to improve classifier.

\section*{Acknowledgments}
This material is based upon work supported by the National Science Foundation  under  Grants  OAC-1934766, CCF-2048223, and CCF-2029044, and the Power System Engineering Research Center (PSERC) under projects S-87.

\balance
\bibliographystyle{IEEEtran}
\bibliography{References.bib}

\begin{IEEEbiography}[{\includegraphics[width=1in,height=1.25in,clip,keepaspectratio]{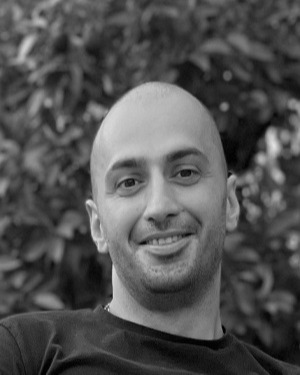}}]{Nima Taghipourbazargani}
received his B.Sc. and M.Sc. degrees in Electrical Engineering from Shahid Beheshti University
and K.N.Toosi University of Technology, Tehran, Iran in 2015, and 2018, respectively. He is currently pursuing the Ph.D. degree with the School of Electrical, Computer and Energy Engineering, Arizona State University. His research interests include application of machine learning and data science in power systems as well as resilience in critical infrastructure networks.
\end{IEEEbiography}

\begin{IEEEbiography}[{\includegraphics[width=1in,height=1.25in,clip,keepaspectratio]{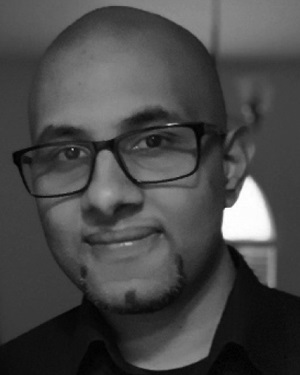}}]{Gautam Dasarathy}
Gautam Dasarathy (Senior Member, IEEE) received the bachelor’s degree in electronics and communication engineering from VIT University in 2008 and the master’s and doctorate degrees in electrical engineering from the University of Wisconsin–Madison in 2010 and 2014, respectively. He held post-doctoral positions at Rice University and Carnegie Mellon University. He is currently an Assistant Professor with the School of Electrical, Computer and Energy Engineering, Arizona State University. His research interests span topics in machine learning, statistics, signal processing and networked systems, and information theory. He was the recipient of the CAREER Award from the National Science Foundation (NSF) in 2021. He was also the winner of the 2022 Distinguished Alumnus Award from VIT University, India.
\end{IEEEbiography}

\begin{IEEEbiography}[{\includegraphics[width=1in,height=1.25in,clip,keepaspectratio]{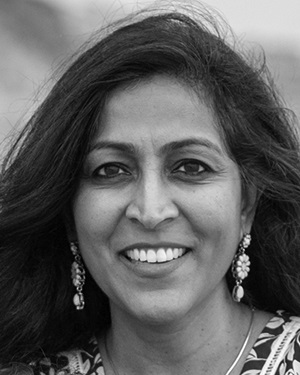}}]{Lalitha Sankar}
Lalitha Sankar (Senior Member, IEEE) received the B.Tech. degree from the Indian Institute of Technology, Mumbai, the M.S. degree from the University of Maryland, and the Ph.D. degree from Rutgers University. She is currently an Associate Professor with the School of Electrical, Computer, and Energy Engineering, Arizona State University. She currently leads an NSF HDR institute on data science for electric grid operations. Her research interests include applying information theory and data science to study reliable, responsible, and privacy-protected machine learning as well as cyber security and resilience in critical infrastructure networks. She received the National Science Foundation CAREER Award in 2014, the IEEE Globecom 2011 Best Paper Award for her work on privacy of side-information in multi-user data systems, and the Academic Excellence Award from Rutgers University in 2008.
\end{IEEEbiography}

\begin{IEEEbiography}[{\includegraphics[width=1in,height=1.25in,clip,keepaspectratio]{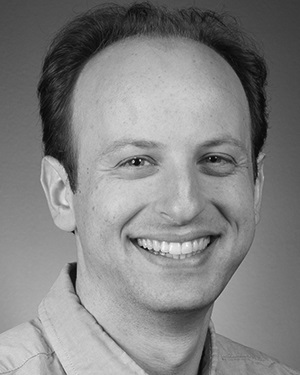}}]{Oliver Kosut}
Oliver Kosut (Member, IEEE) received the B.S. degree in electrical engineering and mathematics from the Massachusetts Institute of Technology, Cambridge, MA, USA, in 2004, and the Ph.D. degree in electrical and computer engineering from Cornell University, Ithaca, NY, USA, in 2010. He was a Post-Doctoral Research Associate with the Laboratory for Information and Decision Systems, MIT, from 2010 to 2012. Since 2012, he has been a Faculty Member with the School of Electrical, Computer and Energy Engineering, Arizona State University, Tempe, AZ, USA, where he is currently an Associate Professor. His research interests include information theory, particularly with applications to security and machine learning, and power systems. Professor Kosut received the NSF CAREER Award in 2015.
\end{IEEEbiography}

\end{document}